\documentclass[prl,a4paper,twocolumn,noshowpacs,floatfix,groupedaddress,amsmath,amsfonts,amssymb,preprintnumbers,nofootinbib,citeautoscript]{revtex4}
\pdfoutput=1
\usepackage[T1]{fontenc}\usepackage[latin1]{inputenc}
\usepackage{dcolumn,graphicx,color,booktabs,wrapfig,microtype,afterpage} \graphicspath{{Figures/}}
\usepackage[charter,greekuppercase=italicized]{mathdesign}\usepackage{sidecap}
\renewcommand{\figurename}{\bf\textsf{Figure}\sffamily}
\renewcommand{\tablename}{Table}
\makeatletter\renewcommand{\fnum@figure}[1]{\textbf{\sffamily\figurename~\thefigure~|\,}}\makeatother
\makeatletter\renewcommand{\fnum@table}[1]{\textbf{\sffamily\tablename~\thetable~|\,}}\makeatother

\newcount\hh \newcount\mm
\hh=\time \divide\hh by 60
\mm=\hh \multiply\mm by 60 \mm=-\mm
\advance\mm by \time
\def\now{\number\hh:\ifnum\mm<10{}0\fi\number\mm}

\definecolor{NatureBlue}{rgb}{0.012,0.3,0.63}
\definecolor{NiceOrange}{rgb}{0.85,0.42,0.21}
\usepackage[colorlinks,plainpages=false,linkcolor=NatureBlue,urlcolor=NatureBlue,citecolor=NatureBlue,pdfpagemode=UseNone,pdfstartview=FitBH]{hyperref}

\makeatletter
\newcommand{\bibstyle@supplement}{\bibpunct[, ]{[S}{]}{;}{n}{,}{,}%
    \gdef\bibnumfmt##1{[S##1]}}
\makeatother

\begin{document}


\title{\flushleft\fontsize{18pt}{20pt}\selectfont\sffamily \textcolor{NiceOrange}{Magnon spectrum of the helimagnetic insulator Cu$_2$OSeO$_3$}}

\author{\sffamily P.\hspace{0.6ex}Y.\hspace{0.6ex}Portnichenko$^1$, J.~Romh\'anyi$^2$, Y.\hspace{0.6ex}A.\hspace{0.6ex}Onykiienko$^1$, A.\hspace{0.6ex}Henschel$^3$, M.~Schmidt$^3$, A.\hspace{0.6ex}S.\hspace{0.6ex}Cameron$^1$, M.\hspace{0.6ex}A.\hspace{0.6ex}Surmach$^1$, J.~A.~Lim$^{1}$, J.\,T.\hspace{0.6ex}Park$^5$, A.\hspace{0.6ex}Schneidewind$^4$, D.~L.~Abernathy$^6$, H.~Rosner$^3$, Jeroen van den Brink$^7$, D.~S.~Inosov$^{1,\,\ast^{\mathstrut}}$\smallskip}
\affiliation{\flushleft\mbox{\sffamily $^1$\hspace{0.5pt}Institut f\"ur Festk\"orperphysik, TU Dresden, Helmholtzstra{\ss}e 10, D-01069 Dresden, Germany.}
\mbox{\sffamily $^2$\hspace{0.5pt}Max Planck Institute for Solid State Research, Heisenbergstra{\ss}e 1, D-70569 Stuttgart, Germany.}
\mbox{\sffamily $^3$\hspace{0.5pt}Max Planck Institute for Chemical Physics of Solids, N\"othnitzer Stra{\ss}e 40, D-01187 Dresden, Germany.}
\mbox{\sffamily $^4$\hspace{0.5pt}J\"ulich Center for Neutron Science (JCNS), Forschungszentrum J\"ulich GmbH, Outstation at Heinz Maier\,--\,Leibnitz} \mbox{$\phantom{^4}$\hspace{0.5pt}\sffamily Zentrum~(MLZ), Lichtenbergstra{\ss}e 1, D-85747 Garching, Germany.}
\mbox{\sffamily $^5$\hspace{0.5pt}Heinz Maier-Leibnitz Zentrum (MLZ), TU M{\"u}nchen, Lichtenbergstra{\ss}e 1, D-85747 Garching, Germany.}
\mbox{\sffamily $^6$\hspace{0.5pt}Quantum Condensed Matter Division, Oak Ridge National Laboratory (ORNL), Oak Ridge, TN 37831, USA.}
\mbox{\sffamily $^7$\hspace{0.5pt}Leibniz Institute for Solid State and Materials Research, IFW Dresden, Helmholtzstra{\ss}e 20, D-01069, Germany.}
\vspace*{3pt}}

\begin{abstract}\citestyle{nature}
\parfillskip=0pt\relax\fontsize{9pt}{11pt}\selectfont
\noindent\textbf{Complex low-temperature ordered states in chiral magnets are typically governed by a competition between multiple magnetic interactions. The chiral-lattice multiferroic Cu$_2$OSeO$_3$ became the first insulating helimagnetic material in which a long-range order of topologically stable spin vortices known as skyrmions was established. Here we employ state-of-the-art inelastic neutron scattering (INS) to comprehend the full three-dimensional spin excitation spectrum of Cu$_2$OSeO$_3$ over a broad range of energies. Distinct types of high- and low-energy dispersive magnon modes separated by an extensive energy gap are observed in excellent agreement with the previously suggested microscopic theory based on a model of entangled Cu$_4$ tetrahedra. The comparison of our INS data with model spin-dynamical calculations based on these theoretical proposals enables an accurate quantitative verification of the fundamental magnetic interactions in Cu$_2$OSeO$_3$ that are essential for understanding its abundant low-temperature magnetically ordered phases.\hfill\vspace{-3pt}}
\end{abstract}

\keywords{multiferroics, magnon excitations, inelastic neutron scattering}
\pacs{75.30.Ds, 75.85.+t, 78.70.Nx\vspace{-0.7em}}

\pagestyle{plain}
\makeatletter
\renewcommand{\@oddfoot}{\hfill\bf\scriptsize\textsf{\thepage}}
\renewcommand{\@evenfoot}{\bf\scriptsize\textsf{\thepage}\hfill}
\makeatother

\citestyle{nature}
\maketitle

\makeatletter\immediate\write\@auxout{\string\bibstyle{my-nature}}\makeatother
\renewcommand\bibsection{\section*{\sffamily\bfseries\footnotesize References\vspace{-10pt}\hfill~}}

\noindent
\textbf{Introduction.} Chiral magnets form a broad class of magnetic materials in which long-range dipole interactions, magnetic frustration, or relativistic Dzyaloshinskii-Moriya (DM) interactions twist the initially ferromagnetic spin arrangement, thus leading to the formation of noncollinear incommensurate helical structures with a broken space-inversion symmetry \cite{TokuraSeki10, Braun12, Kimura12, NagaosaTokura13}.\! Helical magnetic states occur in a very broad range of compounds including metals and alloys, semiconductors, and multiferroics. From the fundamental point of view, the significance of such materials is due to the richness of their possible magnetic structures as compared to ordinary (commensurate and collinear) magnets. Perhaps the most prominent example is given by the topologically nontrivial long-range ordered states known as skyrmion lattices \cite{BogdanovYablonsky89, RosslerBogdanov06, MuhlbauerBinz09}\!.

The interest in the multiferroic ferrimagnet Cu$_2$OSeO$_3$ with a chiral crystal structure has been intensified in recent years after the discovery of a skyrmion-lattice phase in this system \cite{SekiYu12}, thus making it the first known insulator that exhibits a skyrmion order. It was shown soon afterwards that this skyrmion lattice can be manipulated by the application of either magnetic \cite{SekiKim12} or electric \cite{WhiteLevatic12} field, and more recently also by chemical doping \cite{WuWei15}, which indicates a delicate balance of the magnetic exchange and DM interactions with the spin anisotropy effects. Understanding the complex magnetic phase diagram of Cu$_2$OSeO$_3$ therefore requires detailed knowledge of the spin Hamiltonian with precise quantitative estimates of all interaction parameters, which can be obtained from measurements of the spin-excitation spectrum. Microscopic theoretical models that were recently proposed for the description of spin arrangements in Cu$_2$OSeO$_3$ include five magnetic exchange integrals and five anisotropic DM couplings between neighbouring $S=\frac{1}{2}$ copper spins \cite{YangLi12, JansonRousochatzakis14, RomhanyiBrink14, OzerovRomhanyi14, ChizhikovDmitrienko15}, yet their experimental verification was so far limited to thermodynamic data \cite{JansonRousochatzakis14}, tera\-hertz electron spin resonance (ESR) \cite{OzerovRomhanyi14}, far-infrared \cite{MillerXu10} and Raman spectroscopy \cite{GnezdilovLamonova10} that can only probe zone-center excitations in reciprocal space. Here we present the results of inelastic neutron scattering (INS) measurements that reveal the complete picture of magnetic excitations in Cu$_2$OSeO$_3$ accessible to modern neutron spectroscopy over the whole Brillouin zone and over a broad range of energies and demonstrate good quantitative agreement with spin-dynamical model \mbox{calculations both in}

\onecolumngrid\vspace{5pt}

\begin{figure*}[h]
\begin{minipage}{0.61\textwidth}\hspace*{-7pt}
\includegraphics[width=\textwidth]{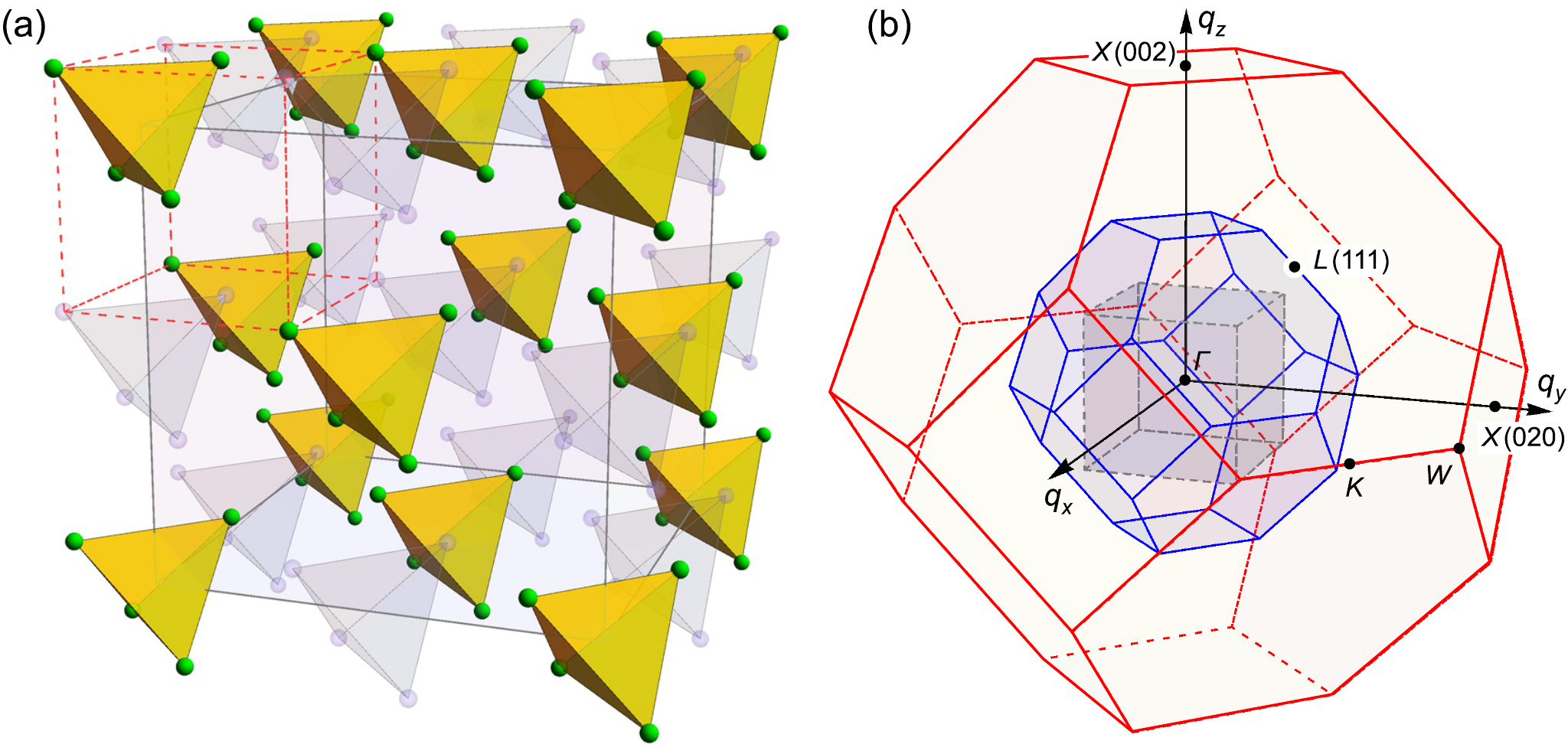}
\end{minipage}\hfill
\begin{minipage}{0.36\textwidth}
\caption{\textbf{Simplified structure of the magnetic Cu sublattice in Cu$_\text{2}$OSeO$_\text{3}$ and Brillouin zone unfolding.} \textbf{a},~The sublattice of Cu atoms approximated by an fcc arrangement of identical tetrahedral Cu$_4$ clusters, the corresponding unit cell shown with grey lines. Introduction of imaginary tetrahedra (faded color) within the voids allows to introduce a twice smaller structural unit cell (dashed red lines) that corresponds to a half-filled fcc lattice of Cu atoms. \textbf{b},~Brillouin zones that correspond to the original crystallographic simple-cubic unit cell with the lattice parameter $a$ (inner grey cube), the fcc unit cell with lattice parameter $a$ (smaller truncated octahedron, blue lines), and the fcc unit cell with lattice parameter $a/2$ (larger truncated octahedron, red lines).\label{Fig:Structure}}
\end{minipage}\hspace*{-2.5pt}
\end{figure*}\clearpage

\begin{figure*}[t!]\vspace{-5pt}
\includegraphics[width=0.76\textwidth]{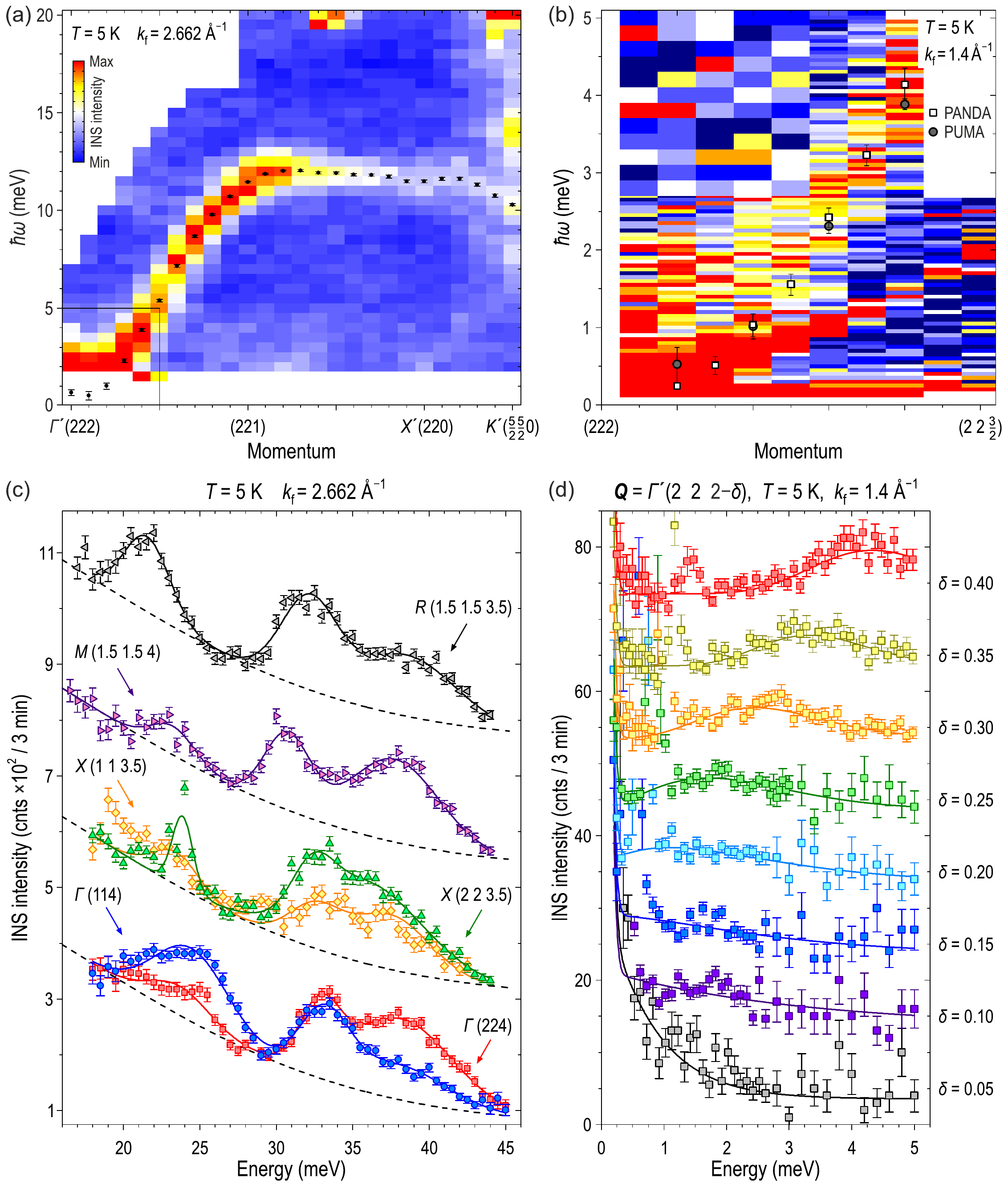}
\caption{\textbf{Magnon excitations in Cu$_\text{2}$OSeO$_\text{3}$ as measured by triple-axis neutron spectroscopy in the low-temperature spin-spiral state (\textit{T}~=~5\,K).} \textbf{a},~Dispersion of the low-energy ferromagnon branch along the $(00H)$ direction mapped out with the thermal-neutron spectrometer PUMA. Data points represent peak positions that resulted from fitting the energy profiles for every momentum. \textbf{b},~The lower part of the same dispersion within the region shown by a black rectangle in panel (a), measured with higher resolution at the cold-neutron spectrometer PANDA. The magnon corresponds to the higher-energy broadened peak (squares), the sharp weaker peaks below it originate from spurious Bragg scattering. The peak positions from the thermal-neutron data in panel (a) are also shown here for comparison (grey circles). \textbf{c}, Raw energy scans across the upper magnon bands measured at several high-symmetry points using the PUMA spectrometer. Here the spectra are grouped and marked according to the small (original crystallographic) cubic Brillouin zone. \textbf{d},~Individual energy cuts from the cold-neutron dataset in panel (b) for different momenta along $(2~2~2\kern.5pt\text{--}{\kern.5pt}\delta)$. The fits shown with solid lines neglect the spurious peaks at the low-energy side of the spectrum.\label{Fig:TAS}\vspace{-7pt}}
\end{figure*}

\twocolumngrid
\noindent terms of magnon dispersions and dynamical structure factors.

\smallskip

\noindent\textbf{Crystal structure and Brillouin zone unfolding.} The cubic copper(II)-oxoselenite Cu$_2$OSeO$_3$ crystallizes in a complex chiral structure with 16 formula units per unit cell (space group $P$2$_1$3, lattice constant $a=8.925$\,\AA) \cite{EffenbergerPertlik86}\!. This structure has been visualized, for instance, in Refs.\,\citenum{SekiYu12} or \citenum{JansonRousochatzakis14}. For the discussion of magnetic properties relevant for our study, it is useful to consider only the magnetic sublattice of copper ions, which can be approximated by a face-centered-cubic (fcc) lattice of identical Cu$_4$ tetrahedra, as shown in Fig.~\ref{Fig:Structure}a. This reduces the volume of the primitive unit cell fourfold with respect to the original simple-cubic structure, resulting in only four Cu atoms per primitive cell. In reciprocal space, this simplified lattice possesses a larger ``unfolded'' Brillouin zone with a volume of \mbox{$4\cdot(2\piup/a)^3$} as shown in Fig.~\ref{Fig:Structure}b with blue lines.

Existing theoretical models \cite{JansonRousochatzakis14, RomhanyiBrink14, OzerovRomhanyi14} suggested that individual Cu$_4$ tetrahedra represent essential magnetic building blocks of Cu$_2$OSeO$_3$. Due to the structural inequivalence of the copper sites within every tetrahedron, its magnetic ground state orients one of the four Cu$^{2+}$ spins antiparallel to three others that are coupled ferromagnetically, resulting in a state with the total spin $S=1$. The interactions between neighboring tetrahedral clusters are at least 2.5 times weaker than intratetrahedron couplings \cite{JansonRousochatzakis14, RomhanyiBrink14, OzerovRomhanyi14} and lead to a ferromagnetic arrangement of their total spins below the Curie temperature, $T_{\rm C}=57$\,K. In addition, weak DM interactions twist the resulting ferrimagnetic state into an incommensurate helical spin structure propagating along the $\langle001\rangle$ direction with a pitch $\lambda_{\rm h}=63\,\text{nm}$, that is 100 times larger than the distance between nearest-neighbor tetrahedra, which corresponds to the propagation vector $k_{\rm h}=0.010$\,\AA$^{-1}$ derived from small-angle neutron scattering (SANS) data \cite{SekiKim12}\!.

As the next step, we note that the same structure can be represented as a half-filled fcc lattice of individual Cu atoms. Indeed, if one supplements the lattice with an equal number of imaginary Cu$_4$ tetrahedra by placing them within the voids of the original structure, as shown in Fig.~\ref{Fig:Structure}a, a twice smaller fcc unit cell with lattice parameter $a/2$ can be introduced (red dashed lines). Its primitive cell contains on average one half of a copper atom and corresponds to the large ``unfolded'' Brillouin zone with a volume of \mbox{$32\cdot(2\piup/a)^3$} that is shown with red lines in Fig.~\ref{Fig:Structure}b. As will be seen in the following, the dynamical structure factor representing the distribution of magnon intensities in Cu$_2$OSeO$_3$ inherits the symmetry of this large Brillouin zone. The described unfolding procedure is therefore useful for reconstructing the irreducible reciprocal-space volume for the presentation of INS data. Throughout this paper, we will denote momentum-space coordinates in reciprocal lattice units corresponding to the crystallographic simple-cubic unit cell (\mbox{1~r.\,l.\hspace{0.5pt}u.~=~$2\piup/a$}), whereas high-symmetry points will be marked in accordance to the large unfolded Brillouin zone as explained in Fig.~\ref{Fig:Structure}b. Equivalent points from higher Brillouin zones will be marked with a prime.

\smallskip

\noindent\textbf{Triple-axis neutron spectroscopy.} We start the presentation of our experimental results with the triple-axis spectroscopy (TAS) data shown in Fig.\,\ref{Fig:TAS}. At low energies, the spectrum is dominated by an intense Goldstone mode with a parabolic dispersion, which emanates from the $\Gamma'(222)$ wave vector and has the highest intensity at this point, as can be best seen from the thermal-neutron data in Fig.\,\ref{Fig:TAS}a. Because this point corresponds to the center of the large unfolded Brillouin zone, we can associate this low-energy mode with a ferromagnetic spin-wave branch anticipated from the theory for the collinear magnetically ordered state \cite{RomhanyiBrink14}\!. Upon moving away from the zone center along the $(001)$ direction, its dispersion reaches a maximum of about 12\,meV, while its intensity is reduced continuously towards the unfolded-zone boundary. Much weaker replicas of the same ferromagnon branch can be also recognized at other points with integer coordinates, like $(221)$ or $(220)$, as they coincide with zone centers of the original crystallographic Brillouin zone but not the unfolded one. We have fitted individual energy scans in Fig.\,\ref{Fig:TAS}a with Gaussian peak profiles to extract the measured magnon dispersion shown as black dots.

For a closer examination of the low-energy part of the ferromagnon branch around the $\Gamma'(222)$ point, we also performed TAS measurements at a cold-neutron spectrometer providing higher energy resolution, as shown in Figs.\,\ref{Fig:TAS}b and \ref{Fig:TAS}d. Here we see a parabolically dispersing ferromagnetic spin wave branch (squares) in good agreement with the dispersion obtained from the thermal-neutron measurements (grey circles). The data are additionally contaminated with spurious Bragg scattering that appears as a sharp straight line below the magnon peak in Fig.\,\ref{Fig:TAS}b or as a strong low-energy peak in \ref{Fig:TAS}d. While fitting the experimental data to extract the magnon dispersion (solid lines in \ref{Fig:TAS}d), this spurious peak had to be masked out. As a result, the ferromagnon mode appears to be gapless within our experimental resolution, which agrees with the earlier microwave absorption measurements, where the spin gap value of only 3~GHz (12~$\mu$eV) has been reported \cite{KobetsDergachev10}\!. By fitting the experimental dispersion at low energies with a parabola, $\hslash\omega=Dq^2$, where $q$ is the momentum transfer along $(001)$ measured relative to the $(222)$ structural Bragg reflection, we evaluated the spin-wave ``stiffness'' $D=52.6$\,meV$\cdot$\AA$^2$, which is comparable to that in the prototypical metallic skyrmion compound MnSi \cite{IshikawaShirane77}\!.

The magnetic Goldstone mode is significantly broadened in both energy and momentum and has a large intrinsic width that is considerably exceeding the instrumental resolution of $\sim$\,0.14\,meV. This can be naturally explained as a result of the incommensurability of the spin-spiral structure and its deviation from the collinear ferrimagnetic order. In the helimagnetic state, the low-energy magnetic excitations are expected to split into so-called helimagnons emanating from the 1$^\text{st}$-, 2$^\text{nd}$-, and higher-order magnetic Bragg reflections \cite{JanoschekBernlochner10, KuglerBrandl15, SchwarzeWaizner15}\!. The incommensurability parameter $k_{\rm h}$ is too small in our case to be resolved with a conventional triple-axis spectrometer. The typical helimagnon energy scale in Cu$_2$OSeO$_3$ can be estimated as $Dk_{\rm h}^2\approx10.5$\,$\mu$eV, that is an order of magnitude lower than in MnSi and also lies well below the energy resolution of the instrument. As a result, multiple unresolved helimagnon branches merge into a single broadened peak as seen in Fig.\,\ref{Fig:TAS}d.

\begin{figure}[t]\vspace{3pt}
\includegraphics[width=\columnwidth]{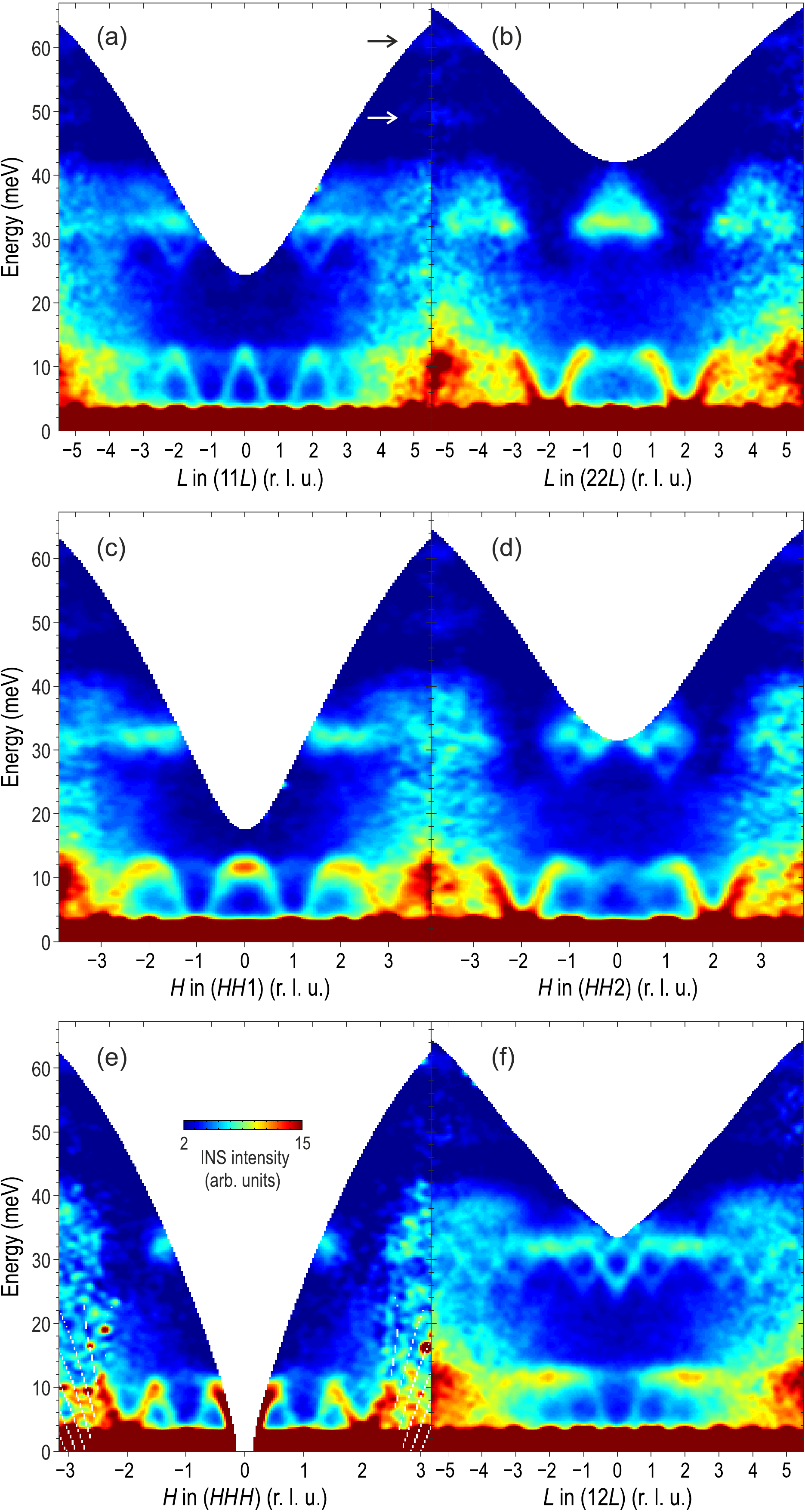}\vspace{2pt}
\caption{\textbf{Representative energy-momentum cuts through the TOF data along the following high-symmetry directions:} \textbf{a},~$(11L)$; \textbf{b},~$(22L)$; \textbf{c},~$(HH1)$; \textbf{d},~$(HH2)$; \textbf{e},~$(HHH)$; \textbf{f},~$(12L)$. All the presented data were measured at the base temperature of 4.5\,K. Black and white arrows mark two high-energy optical phonon modes (see text).\vspace{-7pt}\label{Fig:TOF_QE}}
\end{figure}

According to the theoretical calculations\cite{RomhanyiBrink14}\!, a higher-energy band of dispersive magnon excitations is expected between approximately 25 and 40~meV. In Fig.\,\ref{Fig:TAS}c, we show several spectra measured in this energy range at a number of high-symmetry points that reveal at least three distinct peaks. Here we have grouped the spectra according to their location in the original crystallographic Brillouin zone (grey cube in Fig.\,\ref{Fig:Structure}b) to emphasize that the peak intensities are strongly affected by the dynamical structure factors, whereas their positions in energy in different Brillouin zones remain essentially the same for equivalent wave vectors. The theory anticipates that these high-energy modes represent an intricate tangle of multiple magnon branches, so that several of them may contribute to every experimentally observed peak. A direct comparison with the existing spin-dynamical calculations therefore appears difficult with the limited TAS data and requires a complete mapping of the momentum-energy space using time-of-flight (TOF) neutron spectroscopy that we present below.

\begin{figure*}[t!]\vspace{-2pt}
\includegraphics[width=\textwidth]{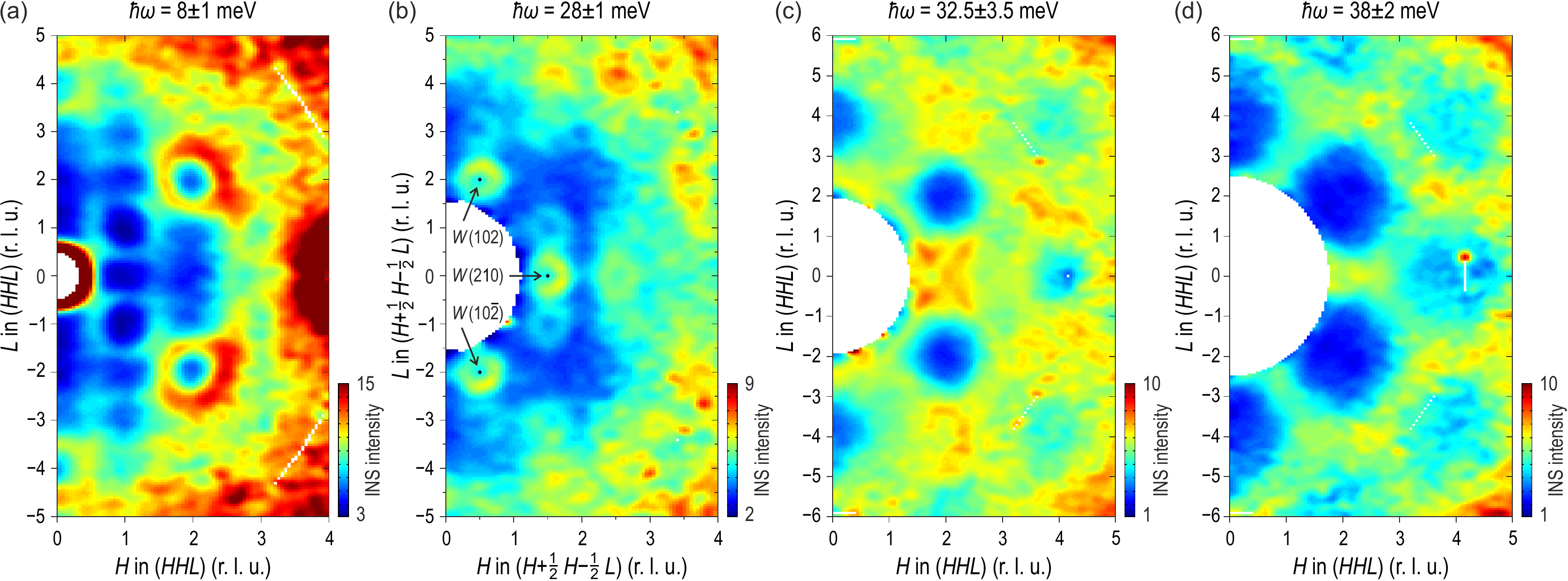}
\caption{\textbf{Constant-energy slices through the TOF data parallel to the $(HHL)$ plane.} \textbf{a},~Along the $(HHL)$ plane at $\hslash\omega=8\pm1$\,meV cutting through the low-energy magnon band with intense rings of scattering from the ferromagnon modes around $\Gamma$ points; \textbf{b},~Along the offset $(H\!-\!\frac{1}{2}~H\!+\!\frac{1}{2}~L)$ plane at $\hslash\omega=28\pm1$\,meV through the bottom part of the upper magnon band, showing rings of intensity around $W$ points; \textbf{c,\,d},~Along the $(HHL)$ plane through the upper magnon band at $\hslash\omega=32.5\pm3.5$\,meV and $38\pm2$\,meV, respectively. In all panels, the data were integrated within $\pm0.15$~r.\,l.\hspace{0.5pt}u. from the indicated plane along the $(1\overline{1}0)$ direction that is orthogonal to the plane of the figure. Corresponding energy integration ranges given above each panel are also shown in Fig.\,\ref{Fig:TOF_Symmetrized}a with vertical bars.\label{Fig:TOF_QxQy}\vspace{-1pt}}
\end{figure*}

\smallskip

\noindent\textbf{Time-of-flight neutron spectroscopy.} The benefit of TOF neutron scattering is that it provides access to the whole 4-dimensional energy-momentum space $(\hslash\omega,\mathbf{Q})$ in a single measurement, which is particularly useful for mapping out complex magnetic dispersions that persist over the whole Brillouin zone. The data can be then analyzed to extract any arbitrary one- or two-dimensional cut from this dataset as we describe in the Methods. Fig.~\ref{Fig:TOF_QE} presents several typical energy-momentum cuts taken along different high-symmetry lines parallel to the $(001)$, $(110)$, and $(111)$ crystallographic axes. One clearly sees both the low-energy ferromagnon band that is most intense around $\Gamma'(222)$ and the intertwined higher-energy dispersive magnon modes between 25 and 40~meV whose intensity anticorrelates with that of the low-energy modes: It vanishes near $\Gamma'(222)$ yet is maximized near the $X'(220)$ point at the unfolded-zone boundary, where the low-energy modes are weak. The low- and high-energy modes are separated by a broad energy gap in the range $\sim$13\,--\,25\,meV. In addition, at much higher energies one can see two relatively weak flat modes centered around 49 and 61\,meV (marked with arrows in Fig.~\ref{Fig:TOF_QE}a). Although the theory predicts a flat magnetic mode around 54\,meV \cite{RomhanyiBrink14}\!, the fact that their intensity monotonically increases towards higher $|\mathbf{Q}|$ speaks rather in favor of optical phonons. It is likely that the 54\,meV magnetic mode is far too weak compared to the phonon peaks and therefore cannot be clearly seen in our data. As the data in Fig.\,\ref{Fig:TOF_QE} extend over several equivalent Brillouin zones in momentum space, one can see an increase of the nonmagnetic background and a simultaneous decrease of the magnetic signal towards higher $|\mathbf{Q}|$, resulting in a rapid reduction of the signal-to-background ratio. Therefore, in the following we restrict our analysis mainly to the irreducible part of the reciprocal space corresponding to shortest possible wave vectors between $\Gamma(000)$ and $\Gamma'(222)$.

It is interesting to note how the hierarchical structure of Brillouin zones introduced in Fig.\,\ref{Fig:Structure} is directly reflected in the magnon intensities. Along the $(HHH)$ direction in Fig.\,\ref{Fig:TOF_QE}e, the strongest spin-wave modes are seen around centers of the large unfolded zones, $\Gamma(000)$ and $\Gamma'(222)$, while a somewhat weaker mode appears at the $L(111)$ point that coincides with the zone center for the smaller fcc Brillouin zone (shown in blue in Fig.\,\ref{Fig:Structure}b). Other points with integer coordinates that are centers of the smallest simple-cubic zones but of none of the larger ones, such as $(110)$ or $(001)$, contain only much weaker replica bands that are barely seen in the data (e.g. Fig.\,\ref{Fig:TOF_QE}a,c or Fig.\,\ref{Fig:TAS}a). This leads to the appearance of rather unusual features in certain cuts through the spectrum, such as the one at the $(001)$ point in the middle of Fig.\,\ref{Fig:TOF_QE}c, which visually resembles the bottom part of the famous hourglass-like dispersion in some transition-metal oxides \cite{HinkovBourges07, BoothroydBabkevich11, DreesLi14}\!. However, here its intense top part stems from the nearest unfolded zone center $\Gamma(000)$ that lies outside of the image plane, while the weaker downward-dispersing branches originate from the $L$ points at $(11\overline{1})$ and $(111)$, producing such a peculiar shape. The situation is even more complicated at higher energies, where the strongest well-resolved downward-pointing modes with a bottom around 26\,meV are located at $W$ points of the large unfolded zone, such as $(102)$, $(210)$ or $(320)$, with weaker replicas shifted by the $(111)$ or equivalent vector. Interestingly, without Brillouin zone unfolding such a particularity of the $W$ points would not at all be obvious without explicit calculations of the dynamical structure factors, as they are all equivalent to the zone center in the original crystallographic Brillouin zone notation.\enlargethispage{5pt}

The pattern of INS intensity in momentum space is visualized by the constant-energy slices presented in Fig.\,\ref{Fig:TOF_QxQy}. All of them are oriented parallel to the $(HHL)$ plane, which was horizontal in our experimental geometry. The low-energy cut integrated around $8\pm1$\,meV (Fig.\,\ref{Fig:TOF_QxQy}a) passes through the origin and contains both the $(222)$ and $(111)$ wave vectors, where rings of scattering from the stronger and weaker magnetic Goldstone modes, respectively, can be clearly seen. The next cut at $28\pm1$\,meV in Fig.\,\ref{Fig:TOF_QxQy}b is shifted by the $(\frac{1}{2}\bar{\frac{1}{2}}0)$ vector in such a way that it crosses the previously mentioned downward-pointing high-energy modes, which appear as small circles around $W$ points. Finally, the last two panels (Fig.\,\ref{Fig:TOF_QxQy}c,d) show integrated intensity in the $(HHL)$ plane at energies corresponding to the middle ($32.5\pm3.5$\,meV) and top ($38\pm2$\,meV) of the upper magnon band, respectively. While individual magnon dispersions are intertwined at these energies and cannot be resolved separately, in both panels we observe deep minima of intensity at the $\Gamma'(222)$, $\Gamma''(400)$, and equivalent wave vectors. They correspond to the suppression of the dynamical structure factor at the center of the unfolded Brillouin zone, where the low-energy ferromagnon mode, on the contrary, is most intense.

\begin{figure*}[t]
\includegraphics[width=\textwidth]{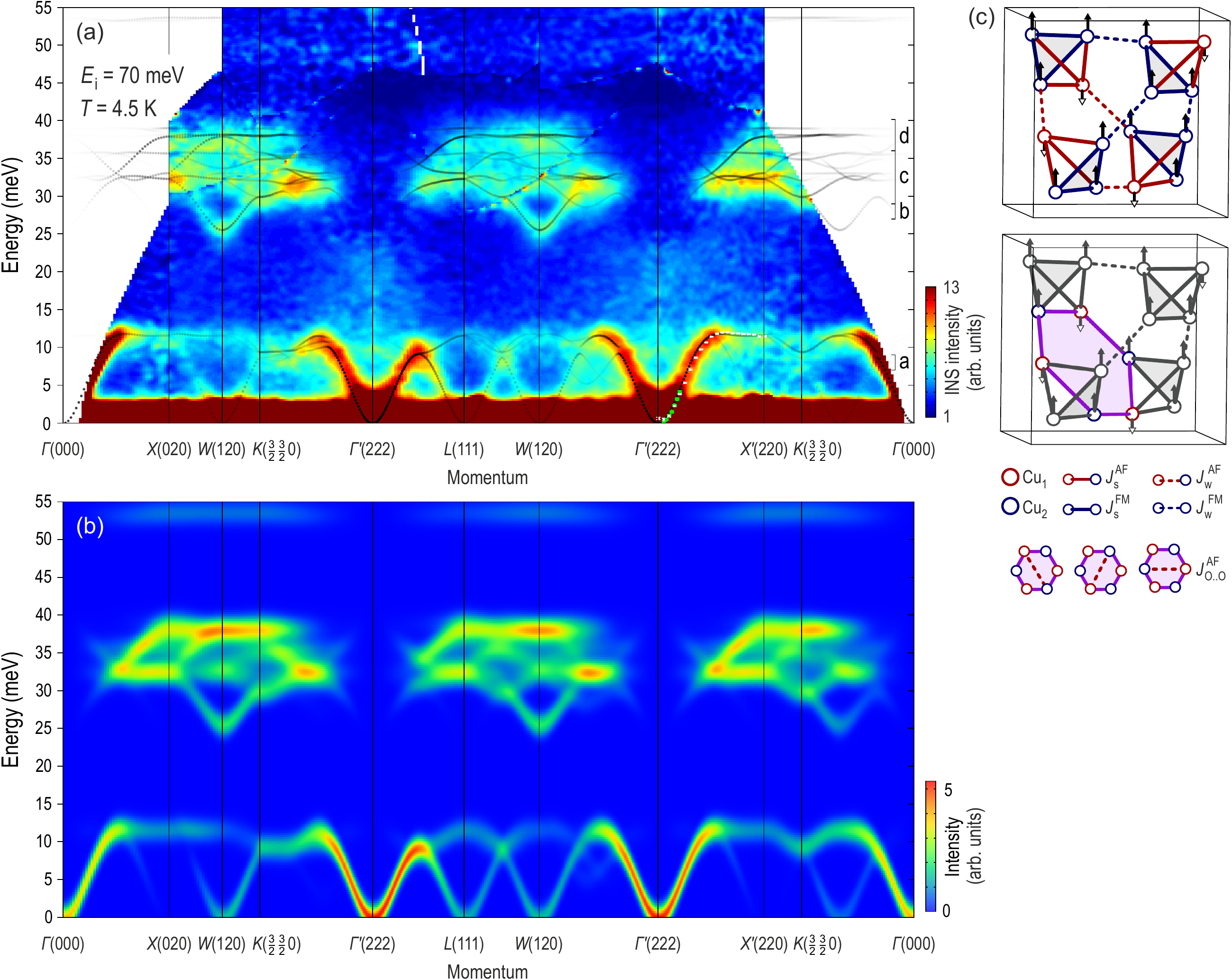}
\caption{\textbf{Symmetrized TOF data along high-symmetry directions.}~\textbf{a},\,Energy-momentum profile of the symmetrized low-temperature TOF data ($T=4.5$\,K) along a polygonal path involving most of the high-symmetry directions in $\mathbf{Q}$-space. The data within every segment have been averaged with all equivalent cuts in the same Brillouin zone as described in the Methods. The peak positions extracted from the fits to the thermal- and cold-neutron TAS data between the $\Gamma'(222)$ and $X'(220)$ points (Fig.\,\ref{Fig:TAS}a,b) are shown here with white and green data points. The vertical bars marked as a\,--\,d on the right-hand side of the panel show energy integration ranges of the corresponding panels in Fig.\,\ref{Fig:TOF_QxQy}. Dotted lines show the calculated dispersions, their transparency reflecting the dynamical structure factors (calculated intensities) as described in the text. \textbf{b},\,Calculated scattering intensity presented as a color map that was obtained by broadening the results of our spin-dynamical calculations with a Gaussian function meant to model the experimental resolution. \textbf{c},\,Exchange paths in Cu$_2$OSeO$_3$. Top: first-neighbor ferromagnetic (blue) and antiferromagnetic (red) couplings. Weak bonds are indicated with dashed lines. Bottom: further-neighbor interaction $J^\text{AF}_\text{O..O}$ realized on the diagonals of hexagons formed by alternating Cu$_1$ and Cu$_2$ sites.\label{Fig:TOF_Symmetrized}}
\end{figure*}

To get a more complete picture of magnetic excitations in Cu$_2$OSeO$_3$ and to compare them with the results of spin-dynamical calculations, we have symmetrized the TOF data (see Methods) and assembled energy-momentum profiles along a polygonal path involving all high-symmetry directions in $\mathbf{Q}$-space into a single map presented in Fig.\,\ref{Fig:TOF_Symmetrized}a. Due to the kinematic constraints, the data in the first Brillouin zone are always limited to lower energies as can be seen in Fig.\,\ref{Fig:TOF_QE}, but have the best signal-to-noise ratio. Therefore, for the best result we also underlayed data from higher Brillouin zones to show the higher-energy part of the spectrum at equivalent wave vectors in the unfolded notation.

\smallskip

\noindent\textbf{Spin-dynamical calculations.} Based on previous theoretical and experimental results \cite{JansonRousochatzakis14, OzerovRomhanyi14, RomhanyiBrink14}\!, we used a tetrahedron-factorized multi-boson method to calculate the magnon spectrum for the collinear ferrimagnetically ordered state. We consider the Heisenberg-type Hamiltonian:\vspace{2pt}
\begin{equation}
\mathcal{H}=\sum_{i,j} J_{ij}\boldsymbol{S}_i\cdot\boldsymbol{S}_j,\vspace{-2.5pt}
\end{equation}
where $J_{ij}$ denote five different kinds of exchange couplings between sites $i$ and $j$ that are explained in Fig.\,\ref{Fig:TOF_Symmetrized}c. There are two strong couplings, a ferromagnetic ($J^\text{FM}_\text{s}$) and an antiferromagnetic ($J^\text{AF}_\text{s}$) one, residing on the Cu$_4$ tetrahedra that form the elementary units of our tetrahedra-factorized multiboson theory. Furthermore, there are weak ferromagnetic ($J^\text{FM}_\text{w}$) and antiferromagnetic ($J^\text{AF}_\text{w}$) first-neighbor interactions connecting the tetrahedra. The last coupling is an antiferromagnetic longer-range interaction ($J_\text{O..O}$) connecting Cu$_1$ and Cu$_2$ across the diagonals of hexagonal loops formed by alternating Cu$_1$ and Cu$_2$ sites. These five interactions are treated on a mean-field level, whereas the antisymmetric DM interactions are neglected for simplicity, as they would affect only the lowest frequency range of the spectrum in the vicinity of the zone center. This same model has been discussed in detail in Ref.~\citenum{RomhanyiBrink14}, for details see also Supplementary Information. Having a complete picture of magnetic excitations in the entire Brillouin zone up to high energies, we can refine the previously established interaction parameters determined from ab-initio calculations \cite{JansonRousochatzakis14} and ESR experiments \cite{OzerovRomhanyi14}\!.\enlargethispage{5pt}

The values for weak interaction parameters were found to be $J_\text{w}^\text{AF}=27$\,K, $J_\text{w}^\text{FM}=-50$\,K and $J^\text{AF}_\text{O..O}=45$\,K, as defined earlier by fitting the magnetization data \cite{JansonRousochatzakis14}\!. These parameters, in fact, provide low-lying modes in agreement with the TOF data as shown in Fig.~\ref{Fig:TOF_Symmetrized}a. Strong coupling constants within the tetrahedra, $J_\text{s}^\text{AF}$ and $J_\text{s}^\text{FM}$, are mainly responsible for the positions of high-energy modes and govern the intra-tetrahedron excitations. We keep $J_\text{s}^\text{AF}=145$\,K as it was determined from fitting the ESR spectrum \cite{OzerovRomhanyi14}\!. Modifying $J_\text{s}^\text{FM}$ to --$170$\,K, we can reproduce the high-energy modes in excellent agreement with experiment, as seen in Fig.\,\ref{Fig:TOF_Symmetrized}a, where the result of the spin-dynamical calculations is shown with dotted lines (see also Supplementary Information).

To fully test our model, we also calculated the scattering cross section. At zero temperature this corresponds to
\begin{equation}
\frac{{\rm d}^2\sigma}{{\rm d}\Omega\,{\rm d}\omega}=\sum_{\alpha,\beta}(\delta_{\alpha\beta}-\hat{k}_{\alpha}\hat{k}_\beta)\,S^{\alpha\beta}(\boldsymbol{k},\omega),\vspace{-5pt}
\end{equation}
where $\alpha,\beta=\{x,y,z\}$, $\hat{k}_\alpha=k_\alpha/|\boldsymbol{k}|$, and $S^{\alpha\beta}(\boldsymbol{k},\omega)$ is the dynamical spin structure factor. Here we used the original crystallographic unit cell and the corresponding cubic Brillouin zone, taking the exact positions of each copper atom within the unit cell into account. The theoretically established scattering intensity plot (artificially broadened to model the experimental resolution) is shown in Fig.\,\ref{Fig:TOF_Symmetrized}b, demonstrating strikingly good agreement with the INS data.

\smallskip

\noindent\textbf{Conclusions.} We have presented the complete overview of spin excitations in Cu$_2$OSeO$_3$ throughout the entire Brillouin zone and over a broad energy range. The data reveal low-energy ferromagnetic Goldstone modes and a higher-energy band of multiple intertwined dispersive spin-wave excitations that are separated by an extensive energy gap. All the observed features can be excellently described by the previously developed theoretical model of interacting Cu$_4$ tetrahedra, given that one of the strong interaction parameters, $J_\text{s}^\text{FM}\!$, is set to a new value of --$170$\,K. The complete list of the dominant parameters for the magnetic Hamiltonian determined here is now able to describe all the available experimental results (magnetization, ESR and INS data) simultaneously. Our model can serve as a starting point for more elaborate low-energy theories that would be able to explain the complex magnetic phase diagram of Cu$_2$OSeO$_3$, including the helimagnetic order and skyrmion-lattice phases, which still represents a challenge of high current interest in solid-state physics.\enlargethispage{2pt}

\smallskip

\vspace{0.8em}
{\footnotesize\noindent
{\sffamily\textbf{Methods.}}\vspace{1ex}

\noindent \textbf{Sample preparation.} The compound Cu$_2$OSeO$_3$ has first been synthesized by a reaction of CuO (Alfa Aesar 99.995\%) and SeO$_2$ (Alfa Aesar 99.999\%) at 300\,$^\circ$C (2 days) and 600\,$^\circ$C (7 days) in evacuated fused silica tubes. Starting from this microcrystalline powder, single crystals of Cu$_2$OSeO$_3$ were then grown by chemical transport reaction using NH$_4$Cl as a transport additive, which decomposes in the vapour phase into ammonia and the transport agent HCl. The reaction was performed in a temperature gradient from 575\,$^\circ$C (source) to 460\,$^\circ$C (sink), and a transport additive concentration of 1~mg/cm$^3$ NH$_4$Cl (Alfa Aesar 99.999\%) \cite{GnezdilovLamonova10}\!. The resulting single crystals with typical sizes of 30\,--\,60\,mm$^3$ were selectively characterized by magnetization, dilatometry, and x-ray diffraction measurements, showing good crystallinity and reproducible behavior of physical properties. The crystals were then coaligned into a single mosaic sample for neutron-scattering measurements with a total mass of approximately 2.5~g using a digital X-ray Laue camera.\smallskip

\noindent \textbf{Triple-axis INS measurements.} The measurements presented in Fig.\,\ref{Fig:TAS} have been performed at the PUMA thermal-neutron spectrometer with a fixed final neutron wave vector $k_{\rm f}=2.662$\,\AA$^{-1}$ and the PANDA cold-neutron spectrometer with $k_{\rm f}=1.4$\,\AA$^{-1}\!$, both at the FRM-II research reactor in Garching, Germany. A pyrolytic graphite or cold beryllium filter, respectively, was installed after the sample to suppress higher-order scattering contamination from the monochromator. The sample was mounted in the $(HHL)$ scattering plane and cooled down with a closed-cycle refrigerator to the base temperature of $\sim$5\,K in both experiments.\smallskip

\noindent \textbf{Time-of-flight measurements.} TOF data in Figs.\,\ref{Fig:TOF_QE}--\ref{Fig:TOF_Symmetrized} were collected using the Wide Angular Range Chopper Spectrometer (ARCS) \cite{AbernathyStone12} at the Spallation Neutron Source, Oak Ridge National Laboratory (ORNL), with the incident neutron energy set to $E_{\rm i}=70$\,meV. Here the base temperature of the sample was at 4.5\,K. The reciprocal space was mapped by performing a full 360$^\circ$ rotation of the sample about the vertical $(1\overline{1}0)$ axis in steps of 1$^\circ\!$, counting about 20~min per angle. The experimental energy resolution measured as the full width at half maximum of the elastic line was $\sim$2.5\,meV.\smallskip

\noindent \textbf{Data analysis.} The TOF data were normalized to a vanadium reference measurement for detector efficiency correction, combined, and transformed into energy-momentum space using the open-source \textsc{Matlab}-based \mbox{\textsc{Horace}} analysis software \cite{Horace}. Due to the high symmetry of the cubic lattice, the TOF dataset could be symmetrized in order to improve statistics in the data and thereby improve the signal-to-noise ratio considerably. After it has been established that the intensity of the INS signal follows the symmetry of the large unfolded Brillouin zone introduced in Fig.\,\ref{Fig:Structure}b, we have assumed this symmetry for the symmetrization. This means that every energy-momentum cut in Fig.\,\ref{Fig:TOF_QE} and every segment of Fig.\,\ref{Fig:TOF_Symmetrized}a has been averaged with all possible equivalent cuts within the same Brillouin zone (cuts from different Brillouin zones with different $|\mathbf{Q}|$ were not averaged). For example, the $LW$ or $LK$ segments forming the radius and apothem of the hexagonal face of the unfolded Brillouin zone boundary, respectively, (Fig.\,\ref{Fig:Structure}b) have 48 equivalent instances along 6 nonparallel directions in the Brillouin zone that can be averaged. As this kind of data analysis is very time consuming and requires significant computational efforts, it still awaits to become the standard practice in neutron-scattering research.\bigskip

\noindent\textbf{\sffamily Acknowledgements.} We would like to thank S.~Zherlitsyn and Y.~Gritsenko for sound velocity measurements that assisted our data interpretation and M.~Rotter for helpful discussions at the start of this project. The work at the TU Dresden was financially supported by the German Research Foundation within the collaborative research center SFB\,1143 and the research training group GRK\,1621. Research at ORNL's Spallation Neutron Source was sponsored by the Scientific User Facilities Division, Office of Basic Energy Sciences, US Department of Energy.\smallskip

\noindent{\textbf{\sffamily Author~contributions.} A.\,H. and M.\,S. have grown single crystals for this study; P.~Y.\,P., A.\,S.\,C., M.\,A.\,S., J.\,A.\,L. and D.\,S.\,I. conducted the INS experiments; J.\,T.~P., A.\,S. and D.\,L.\,A. provided instrument support at different neutron spectrometers; P.~Y.\,P. and Y.~A.\,O. analyzed the neutron data and prepared the figures; J.\,R. and J.\,v.\,d.\,B. provided theory support and performed the spin-dynamical calculations; P.~Y.~P., J.\,R., H.\,R., J.\,v.\,d.\,B. and D.\,S.\,I. interpreted the results; D.\,S.\,I. supervised the project and wrote the manuscript; all authors discussed the results and commented on the paper.}\smallskip

\noindent{\textbf{\sffamily Author information.} The authors declare no competing financial interests. Correspondence and requests for materials should be addressed to D.\,S.\,I. $\langle$\href{mailto:dmytro.inosov@tu-dresden.de}{dmytro.inosov@tu-dresden.de}$\rangle$.}

}

\onecolumngrid\clearpage

\vfill
\clearpage

\renewcommand\thefigure{S\arabic{figure}}
\renewcommand\thetable{S\arabic{table}}
\renewcommand\theequation{S\arabic{equation}}
\renewcommand\bibsection{\section*{\sffamily\bfseries\footnotesize Supplementary References\vspace{-6pt}\hfill~}}

\citestyle{supplement}

\pagestyle{plain}
\makeatletter
\renewcommand{\@oddfoot}{\hfill\bf\scriptsize\textsf{S\thepage}}
\renewcommand{\@evenfoot}{\bf\scriptsize\textsf{S\thepage}\hfill}
\renewcommand{\@oddhead}{P.\hspace{0.6ex}Y.\hspace{0.6ex}Portnichenko \textit{et~al.}\hfill\Large\textsf{\textcolor{NiceOrange}{SUPPLEMENTARY INFORMATION}}}
\renewcommand{\@evenhead}{P.\hspace{0.6ex}Y.\hspace{0.6ex}Portnichenko \textit{et~al.}\Large\textsf{\textcolor{NiceOrange}{SUPPLEMENTARY INFORMATION}}\hfill}
\makeatother
\setcounter{page}{1}\setcounter{figure}{0}\setcounter{table}{0}\setcounter{equation}{0}

\makeatletter\immediate\write\@auxout{\string\bibstyle{my-apsrev}}\makeatother

\onecolumngrid\normalsize

\begin{center}{\vspace*{0.1pt}\Large{Supplementary Information to the Letter\smallskip\\\sl\textbf{``\hspace{1pt}Magnon spectrum of the helimagnetic insulator Cu$_2$OSeO$_3$''}}}\end{center}\bigskip

\smallskip
\noindent\textbf{\it Comparison of the theoretical models.}\vspace*{5pt}

\begin{figure*}[b!]
\hspace*{0.11\textwidth}\includegraphics[width=0.78\textwidth]{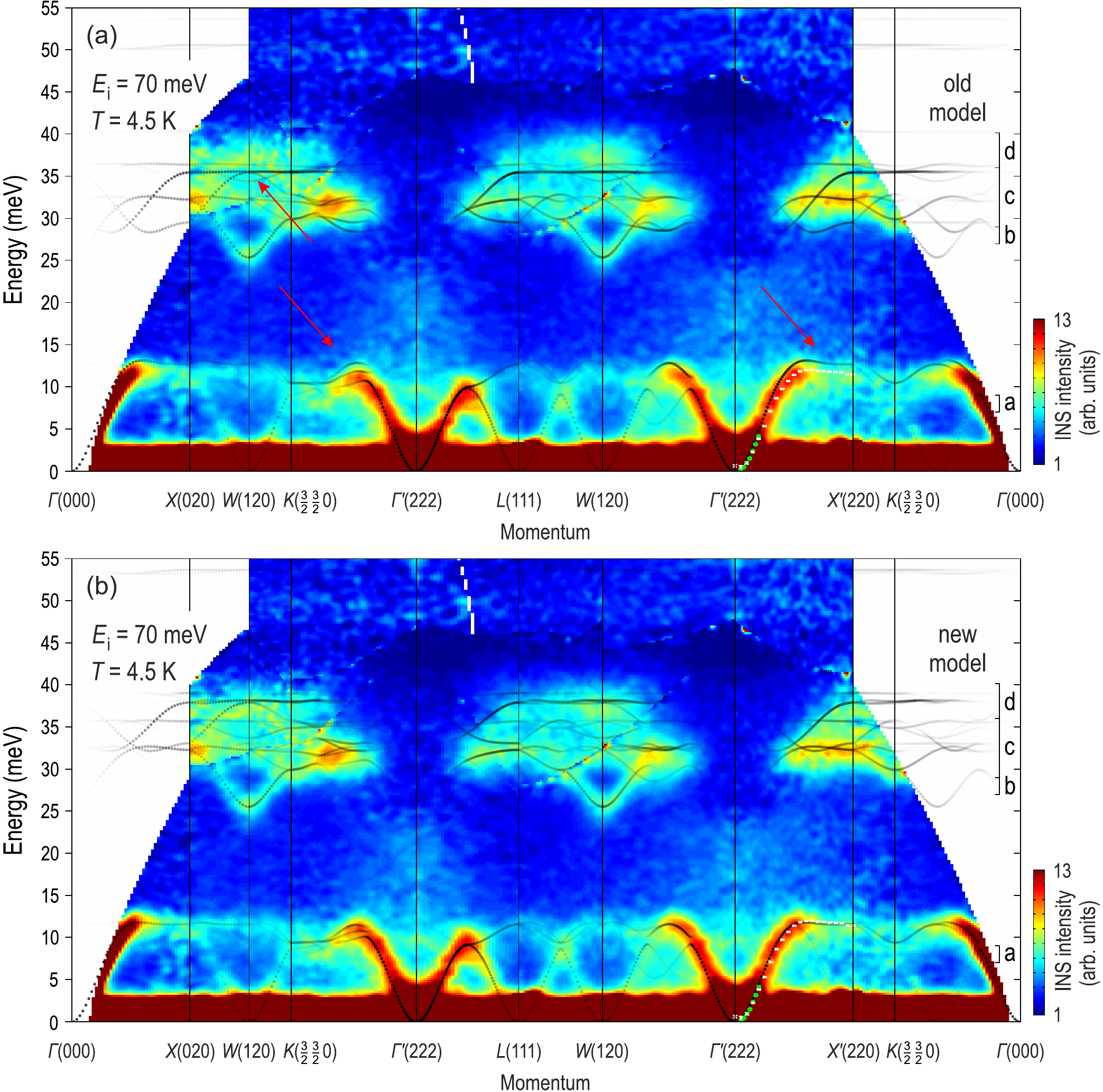}
\caption{The same energy-momentum profile of the symmetrized low-temperature TOF data as shown in Fig.\,\ref{Fig:TOF_Symmetrized}a of the paper, compared with the theoretically calculated dispersions (dotted lines) for two sets of exchange parameters: \textbf{a},~The original set of parameters from Ref.\,\citenum{RomhanyiBrink14}; \textbf{b},~the modified set of parameters presented in our current work, where the strong ferromagnetic exchange interaction $J_\text{s}^\text{FM}$ has been modified to --$170$\,K, resulting in a better match with the experimental data.\label{Fig:TOF_v1_v2}\bigskip}
\end{figure*}

\twocolumngrid
\noindent In Fig.\,\ref{Fig:TOF_v1_v2}, we compare our time-of-flight neutron data with theoretical calculations done using the originally published exchange parameters from Ref.\,\citenum{RomhanyiBrink14} (top) and the modified model with $J_\text{s}^\text{FM} = -170$\,K (bottom) that demonstrates a significantly better agreement with the experimental data, especially in the regions marked with red arrows. In particular, in the high-energy region the old model produced a nearly dispersionless intense feature that coincided with a minimum in the observed inelastic scattering intensity, whereas in the corrected model this magnon branch is shifted towards higher energies.
\onecolumngrid

\twocolumngrid

\begin{figure}[t!]\vspace{-2pt}
\includegraphics[width=\columnwidth]{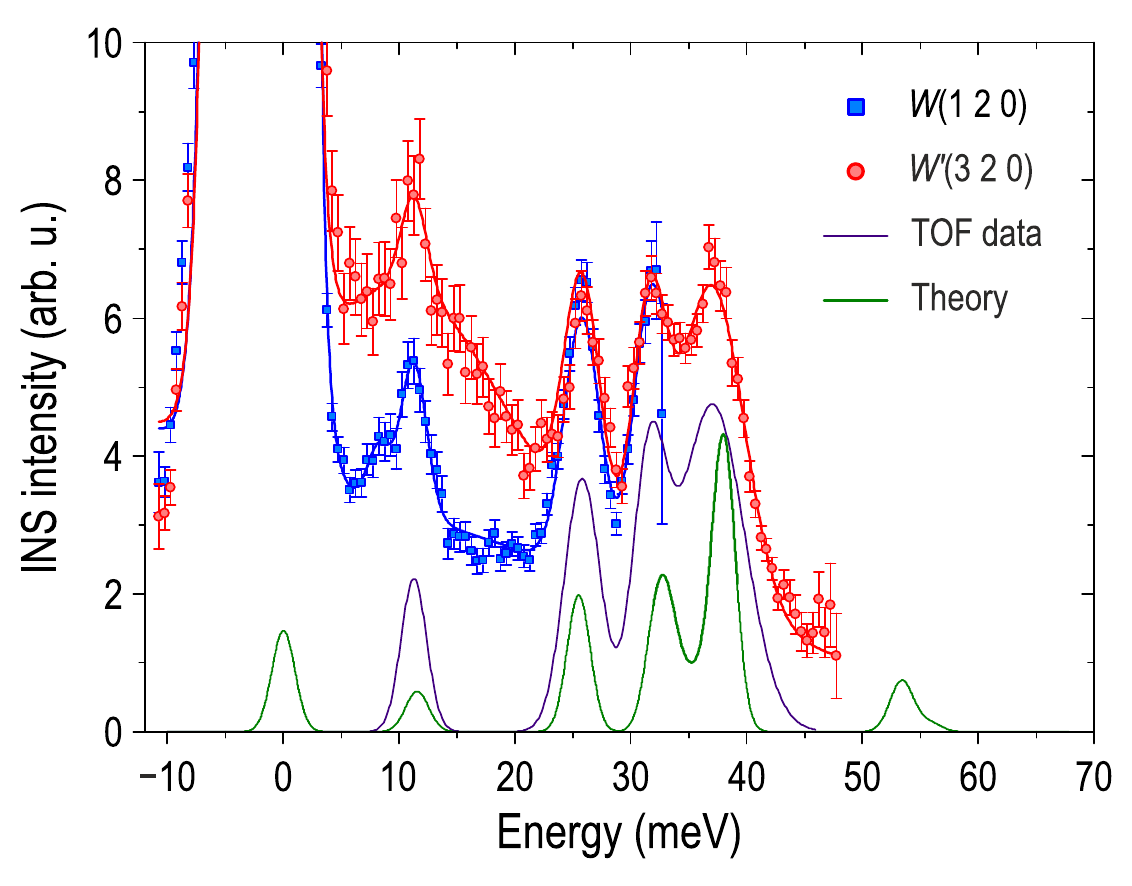}\vspace{-2pt}
\caption{Comparison of the line profiles at the $W(120)$ and $W(320)$ points with the corresponding profiles calculated from the theoretical model after introducing a finite resolution broadening of the magnon peaks. Along with the experimental data (blue and red symbols), we additionally show the result of their fitting with a set of Gaussian peaks (blue solid line) in order to separate them from the energy-dependent background and the model curve obtained from the theory (green line).\label{Fig:W120}}
\end{figure}

In Fig.\,\ref{Fig:W120}, we additionally compare the experimental line profiles at the $W$ points with the corresponding simulated profiles obtained from the calculated magnon dispersions after resolution broadening (as shown in Fig.\,\ref{Fig:TOF_Symmetrized}b in the main text). Because the covered energy range for the $W(120)$ point is only limited to lower energies, we additionally show a cut at the equivalent $W'(320)$ point which provides information about the higher-energy part of the spectrum. Both points were symmetrized with the same points in equivalent Brillouin zones obtained by applying all cyclic permutations of $(HKL)$ with possible sign changes to obtain the best signal-to-noise ratio. After that both datasets were fitted globally with a set of Gaussian peaks to separate the magnetic signal from the energy-dependent background. This magnetic signal is shown as a solid line and compared with the corresponding intensity profile from the theory.\clearpage

\noindent\textbf{\it Multiboson theory for $\text{Cu$_2$OSeO$_3$}$}\vspace*{5pt}

\noindent Having weak and strong interactions suggests that the elementary units are not the bare Cu ions, but the quantum mechanically (QM) entangled tetrahedra formed by four  Cu sites. Therefore, we consider tetrahedra-factorized variational wave function to describe the ground sate of the system and to build upon to introduce the elementary excitations:\vspace{-10pt}
\begin{equation}
|\Psi\rangle=\prod^4_{t=1} |\psi\rangle_t,\vspace{-2pt}
\label{eq:PSI}
\end{equation}
where $|\psi\rangle_t$ is a QM state in the $2^4=16$ dimensional local Hilbert space of the strong tetrahedron $t$. This 16-dimensional Hilbert space can be reduced according to the $C_{3v}$ local symmetry of the distorted tetrahedron (built of one Cu$_1$ and three Cu$_2$ sites) as well as the SU(2) rotational symmetry. Such classification proves to be useful in labelling our states and establishing their mixing without  elaborate calculations. Table~\ref{tab:basis} lists the symmetry classified eigenbasis of a strong tetrahedron.
\begin{table*}[t!]
\caption{The basis of a (strong) tetrahedron classified according to the spin rotational SU(2) and geometrical C$_{3v}$ symmetries.\smallskip\label{tab:basis}}
\begin{tabular}{l@{~~~~}c@{~~~~}l@{~~~~}c@{~~~~}c}
\toprule
irrep & state & notation & $~~~~a^{\dagger}_n$ boson & energy \\
\midrule
\noalign{\smallskip}
{}
& $\frac{1}{2\sqrt{3}}\bigl(|\Downarrow\downarrow\downarrow\uparrow\rangle+|\Downarrow\downarrow\uparrow\downarrow\rangle+|\Downarrow\uparrow\downarrow\downarrow\rangle-3 |\Uparrow\downarrow\downarrow\downarrow\rangle\bigr)$ & $|1,\overline{1}\rangle_{A_1}$ & $a^\dagger_{1,\overline{1},A_1} $ & {}\\
{${\rm A}_1\otimes\,\mathcal{D}^{(1)}$}
& $\frac{1}{\sqrt{6}}\bigl(|\Downarrow\downarrow\uparrow\uparrow\rangle+|\Downarrow\uparrow\downarrow\uparrow\rangle+|\Downarrow\uparrow\uparrow\downarrow\rangle- |\Uparrow\downarrow\downarrow\uparrow\rangle- |\Uparrow\downarrow\uparrow\downarrow\rangle- |\Uparrow\uparrow\downarrow\downarrow\rangle\bigr)$ & $|1,0\rangle_{A_1}$  & $a^\dagger_{1,0,A_1} $ & $E_{1,A_1}=-\frac{5}{4} J_\text{s}^\text{AF}+\frac{3}{4}J_\text{s}^\text{FM}$
\\
{}
& $\frac{1}{2\sqrt{3}}\bigl(3 |\Downarrow\uparrow\uparrow\uparrow\rangle-|\Uparrow\downarrow\uparrow\uparrow\rangle-|\Uparrow\uparrow\downarrow\uparrow\rangle-|\Uparrow\uparrow\uparrow\downarrow\rangle\bigr)$ & $|1,1\rangle_{A_1}$ & $a^\dagger_{1,1,A_1} $ & {}\\
\noalign{\smallskip}
\midrule
\noalign{\smallskip}
{${\rm E}\,\otimes\,\mathcal{D}^{(0)}$}
& $\left.\begin{array}{c}
\frac{1}{2\sqrt{3}}\bigl(2|\Downarrow\downarrow\uparrow\uparrow\rangle-|\Downarrow\uparrow\downarrow\uparrow\rangle-|\Downarrow\uparrow\uparrow\downarrow\rangle- |\Uparrow\downarrow\downarrow\uparrow\rangle- |\Uparrow\downarrow\uparrow\downarrow\rangle+2 |\Uparrow\uparrow\downarrow\downarrow\rangle\bigr)\\
\frac{1}{2}\bigl(|\Downarrow\uparrow\downarrow\uparrow\rangle-|\Downarrow\uparrow\uparrow\downarrow\rangle- |\Uparrow\downarrow\downarrow\uparrow\rangle+ |\Uparrow\downarrow\uparrow\downarrow\rangle\bigr)\end{array}\right\}
$ & $|0,0\rangle_{E}$ & $\begin{array}{c} a^\dagger_{0,0,E_{(1)}} \\ a^\dagger_{0,0,E_{(2)}} \end{array}$ & $E_{0,E}=-\frac{3}{4} J_\text{s}^\text{AF}-\frac{3}{4}J_\text{s}^\text{FM}$
\\
\noalign{\smallskip}
\midrule
\noalign{\smallskip}
{}
& $|\Downarrow\downarrow\downarrow\downarrow\rangle$ & $|2,\overline{2}\rangle_{A_1}$ & $a^\dagger_{2,\overline{2},A_1} $ & {} \\
{}
& $\frac{1}{2}\bigl(|\Downarrow\downarrow\downarrow\uparrow\rangle+|\Downarrow\downarrow\uparrow\downarrow\rangle+|\Downarrow\uparrow\downarrow\downarrow\rangle+ |\Uparrow\downarrow\downarrow\downarrow\rangle\bigr)$ & $|2,\overline{1}\rangle_{A_1}$  & $a^\dagger_{2,\overline{1},A_1} $ & {}\\
{${\rm A}_1\otimes\,\mathcal{D}^{(2)}$}
& $\frac{1}{\sqrt{6}}\bigl(|\Downarrow\downarrow\uparrow\uparrow\rangle+|\Downarrow\uparrow\downarrow\uparrow\rangle+|\Downarrow\uparrow\uparrow\downarrow\rangle+ |\Uparrow\downarrow\downarrow\uparrow\rangle+ |\Uparrow\downarrow\uparrow\downarrow\rangle+ |\Uparrow\uparrow\downarrow\downarrow\rangle\bigr)$ & $|2,0\rangle_{A_1}$ & $a^\dagger_{2,0,A_1} $ & $E_{2,A_1}=\frac{3}{4}J_\text{s}^\text{AF}+\frac{3}{4}J_\text{s}^\text{FM}$\\
{} & $\frac{1}{2}\bigl(|\Downarrow\uparrow\uparrow\uparrow\rangle+|\Uparrow\downarrow\uparrow\uparrow\rangle+|\Uparrow\uparrow\downarrow\uparrow\rangle+|\Uparrow\uparrow\uparrow\downarrow\rangle\bigr)$ & $|2,1\rangle_{A_1}$ & $a^\dagger_{2, 1,A_1} $ & {}\\
{}
& $|\Uparrow\uparrow\uparrow\uparrow\rangle$ & $|2,2\rangle_{A_1}$ & $a^\dagger_{2,2,A_1} $ & {}\\
\noalign{\smallskip}
\midrule
\noalign{\smallskip}
{}
& $\left.\begin{array}{c}
\frac{1}{\sqrt{6}}\bigl(|\Downarrow\downarrow\downarrow\uparrow\rangle+|\Downarrow\downarrow\uparrow\downarrow\rangle-2|\Downarrow\uparrow\downarrow\downarrow\rangle\bigr)\\
\frac{1}{\sqrt{2}}\bigl(|\Downarrow\downarrow\downarrow\uparrow\rangle-|\Downarrow\downarrow\uparrow\downarrow\rangle\bigr)\end{array}\right\}
$ & $|1,\overline{1}\rangle_{E}$ & $\begin{array}{c} a^\dagger_{1,\overline{1},E_{(1)}} \\ a^\dagger_{1,\overline{1},E_{(2)}} \end{array}$ & {}\\
\noalign{\smallskip}
{${\rm E}\,\otimes\,\mathcal{D}^{(1)}$}
& $\left.\begin{array}{c}
\frac{1}{2\sqrt{3}}\bigl(-2|\Downarrow\downarrow\uparrow\uparrow\rangle+|\Downarrow\uparrow\downarrow\uparrow\rangle+|\Downarrow\uparrow\uparrow\downarrow\rangle- |\Uparrow\downarrow\downarrow\uparrow\rangle- |\Uparrow\downarrow\uparrow\downarrow\rangle+2 |\Uparrow\uparrow\downarrow\downarrow\rangle\bigr)\\
\frac{1}{2}\bigl(|\Downarrow\uparrow\downarrow\uparrow\rangle-|\Downarrow\uparrow\uparrow\downarrow\rangle+ |\Uparrow\downarrow\downarrow\uparrow\rangle- |\Uparrow\downarrow\uparrow\downarrow\rangle\bigr)\end{array}\right\}
$ & $|1,0\rangle_{E}$ & $\begin{array}{c} a^\dagger_{1,0,E_{(1)}} \\ a^\dagger_{1,0,E_{(2)}} \end{array}$ & $E_{1,E}=\frac{1}{4}J_\text{s}^\text{AF}-\frac{3}{4}J_\text{s}^\text{FM}$
\\
\noalign{\smallskip}
{}
& $\left.\begin{array}{c}
\frac{1}{\sqrt{6}}\bigl(-2|\Uparrow\downarrow\uparrow\uparrow\rangle+|\Uparrow\uparrow\downarrow\uparrow\rangle+|\Uparrow\uparrow\uparrow\downarrow\rangle\bigr)\\
\frac{1}{\sqrt{2}}\bigl(|\Uparrow\uparrow\downarrow\uparrow\rangle-|\Uparrow\uparrow\uparrow\downarrow\rangle\bigr)\end{array}\right\}
$ & $|1,1\rangle_{E}$ & $\begin{array}{c} a^\dagger_{1,1,E_{(1)}} \\ a^\dagger_{1,1,E_{(2)}} \end{array}$ & {}\\
\noalign{\smallskip}
\bottomrule
\end{tabular}
\end{table*}

The ground state is an $A_1$ triplet separated with a large energy gap of $\sim$280\,K from the rest of the tetrahedron spectrum. A finite weak interaction between tetrahedra mixes the multiplets, i.e. states with different total spin $S$. The local symmetry, however, remains $C_{3v}$, and the $z$ component of the spin remains a good quantum number. Thus the $A_1$ triplet can only mix with the $A_1$ quintet. Furthermore, if we select the $S^z=1$ state as the ground state, purely for sake of simplicity, the state $|1,1,A_1\rangle$ couples solely to $|2,1,A_1\rangle$ in the tetrahedra-factorized picture.
The ground state in the tetrahedra mean-field case reads
\begin{equation}
|\psi\rangle_t=\cos\frac{\alpha}{2}|1,1,A_1\rangle_t+\sin\frac{\alpha}{2}|2,1,A_1\rangle.\vspace{-3pt}
\label{eq:psi_t}
\end{equation}
The variational parameter $\alpha$ controls the length of local spins in the following manner:
\begin{equation}
\langle S^z_1\rangle=-\frac{1}{4}\left(\cos\alpha+\sqrt{3}\sin\alpha\right)\text{~~and~~}\langle S^z_2\rangle=\frac{1}{3}\left(1-\langle S^z_1\rangle\right).
\end{equation}
Note that for any $\alpha$ the total magnetization per tetrahedron remains $\langle S^z_1\rangle+3\langle S^z_2\rangle=1$, i.e. we remain in the one-half plateau phase regardless the value of the variational parameter $\alpha$. In the limit $\alpha=0$ we recover the decoupled case with the local moments $\langle S^z_1\rangle=1/4$ and $\langle S^z_2\rangle=5/12$. In the coupled case for coupling values $J_\text{s}^\text{AF}=145$\,K, $J_\text{s}^\text{FM}=-170$\,K, $J_\text{w}^\text{AF}=27$\,K, $J_\text{w}^\text{FM}=-50$\,K, and $J^\text{AF}_\text{O..O}=45$\,K, the variational parameter takes the value $\alpha=0.38191$ and the local moments become $\langle S^z_1\rangle=-0.39337$ and $\langle S^z_2\rangle=0.464457$.

To include quantum fluctuations and describe the dynamical properties, we use multiboson theory\,---\,a generalizisation of the standard spin-wave approach. As a first step, we introduce the boson operators $a^\dagger_{n,t}$ that create the $n^\text{th}$ state ($n=1..16$) of a tetrahedron $t$ ($t=1..4$). As each tetrahedron can only be in one of its 16 states, the bosons must satisfy the constraint\vspace{-3pt}
\begin{equation}
\sum^{16}_{n=1}a^\dagger_{n,t}a^{\phantom\dagger}_{n,t}=1.\vspace{-3pt}
 \end{equation}

The bosonic representation of the spin operator $S^{\alpha}_j$, where $j$ denotes a given copper site, is too excessive to show here. Nonetheless, some operators acquire a simpler form, such as the intra-tetrahedron Hamiltonian and the total magnetization $S^z=\sum_{i=1-4} S^z_i$, which are diagonal in this basis:
\begin{widetext}
\begin{eqnarray}
\mathcal{H}_{t}\!=\!\!E_{\!1,A_1\!}\!\!\sum_{m=\!\overline{1},0,1\!}\!a^\dagger_{\!1,m,A_1\!}a^{\phantom{\dagger}}_{\!1,m,A_1\!}
\!+\!E_{\!0,E\!}\!\sum_{E=E_1,E_2}\!a^\dagger_{\!0,0,E\!}a^{\phantom{\dagger}}_{\!0,0,E\!}\!+\!E_{\!2,A_1\!}\sum_{\!m=\overline{2},\dots,2\!}a^\dagger_{\!1,m,A_1\!}a^{\phantom{\dagger}}_{\!1,m,A_1\!}
\!+\!E_{\!1,E\!}\!\sum_{\substack{\!m=\overline{1},0,1\!\\E=\!E_1,E_2\!}}a^\dagger_{\!1,m,E\!}a^{\phantom{\dagger}}_{\!1,m,E\!},
\end{eqnarray}
\end{widetext}
\begin{equation}
S^z_t=\sum m~ a^\dagger_{S,m,R}a^{\phantom{\dagger}}_{S,m,R}.
\end{equation}
The spin raising operator of a tetrahedron $t$ can be written as
\begin{eqnarray}
S^+_t=\sum\sqrt{S(S+1)-m(m+1)}a^\dagger_{S,m+1,R}a^{\phantom{\dagger}}_{S,m,R},
\end{eqnarray}
where the sum is over the 16-dimensional tetrahedron basis, and the spin lowering operator is its hermitian conjugate.

After rewriting our spin operators in terms of boson operators $a^{\dagger}_{n,t}$, we transform out the boson that creates the ground state to get a Hamiltonian containing only the excitations. Then we solve this spin-wave Hamiltonian using the Bogoliubov transformation to obtain the spectrum. Our ground state is expressed in Eqs.~\ref{eq:PSI} and~\ref{eq:psi_t}. Let $a^\dagger_{1,t}$ denote the boson that creates $|\psi\rangle_t$ at a tetrahedron $t$. The orthogonal bosons will represent the 15 local excitations on each tetrahedron. Thus alltogether we have $4\times15=60$ excitations. To eliminate  $a^\dagger_{1,t}$ we use the above constraint and perform the following substitution
\begin{equation}
a^\dagger_{1,t}\to1-\frac{1}{2}\sum_{n>1} a^{\dagger}_{n,t}a^{\phantom\dagger}_{n,t}\hspace{1.2ex}\text{and}\hspace{1.2ex} a^{\phantom\dagger}_{1,t}\to1-\frac{1}{2}\sum_{n>1} a^{\dagger}_{n,t}a^{\phantom\dagger}_{n,t}.
\end{equation}
In the linearized multiboson theory we only keep second order boson terms in the resulting Hamiltonian. Therefore, the multiboson Hamiltonian, now containing only the excitation boson operators, has zero-order boson terms, i.e. constant terms $\mathcal{H}^{(0)}$, linear terms $\mathcal{H}^{(1)}$, containing single-boson operators, and two-boson operator terms $\mathcal{H}^{(2)}$. $\mathcal{H}^{(0)}$ corresponds to the tetrahedral mean-field ground state energy, $\mathcal{H}^{(1)}$ is identically zero when the variational wave function minimizes the energy, and the quadratic $\mathcal{H}^{(2)}$ terms describe the boson hopping processes, providing the spectrum.

In particular, $\mathcal{H}^{(2)}$ takes the following form:
\begin{eqnarray}
\mathcal{H}^{(2)}=\frac{1}{2}\left(\begin{array}{c}{\bf A}^\dagger_{\bf k}\\{\bf A}^{\phantom{\dagger}}_{-{\bf k}}\end{array}\right)^T
\left(\begin{array}{cc}
M & N\\
N^\dagger & M\end{array}\right)
\left(\begin{array}{c}{\bf A}^{\phantom{\dagger}}_{{\bf k}}\\{\bf A}^\dagger_{-{\bf k}}\end{array}\right)
\end{eqnarray}
where $M=M^\dagger$ and $N=N^T$ are $60\times 60$ matrices and the vector ${\bf A}^\dagger_{\bf k}$ contains the 60 excitations of the unit cell (15 per tetrahedron), namely 60 bosons $a^{\dagger}_{n,t}$ with $n=2,\hdots,16$ and $t=1,\hdots,4$:
\begin{equation}
{\bf A}^\dagger_{\bf k}=\left( a^\dagger_{{\bf k},2,1},\dots, a^\dagger_{{\bf k},16,4}\right),
\end{equation}

The equation of motion reads
\begin{eqnarray}
i\dot{a}^{\phantom\dagger}_{n,t}({\bf k})=[a^{\phantom\dagger}_{n,t}({\bf k}),\mathcal{H}^{(2)}]=\omega_{n,t}({\bf k}) a^{\phantom\dagger}_{n,t}({\bf k}).
\end{eqnarray}
In order to find the linear combinations of bosons that diagonalize $\mathcal{H}^{(2)}$ we need to solve the eigenvalue problem\vspace{-5pt}
\begin{eqnarray}
\left(\begin{array}{cc}M & N\\-N^\dagger & -M\end{array}\right)^T
\left(\begin{array}{c} {\bf u}_\mu\\{\bf v}_\mu\end{array}\right)=
\omega_\mu\left(\begin{array}{c} {\bf u}_\mu\\{\bf v}_\mu\end{array}\right)\vspace{-7pt}
\label{eq:eigen_eq}
\end{eqnarray}\vspace{1pt}

\noindent The physical eigenvectors (with positive eigenvalues) are associated with the states that diagonalize the Hamiltonian $\mathcal{H}^{(2)}$:\vspace{-10pt}
\begin{eqnarray}
\alpha^\dagger_{\mu,{\bf k}}={\bf u}_{\mu}\cdot {\bf A}^{\dagger}_{\bf k}+{\bf v}_{\mu}\cdot {\bf A}^{\phantom{\dagger}}_{-{\bf k}}
\label{eq:bogoliubov}
\end{eqnarray}
Then $\left[\mathcal{H}^{(2)},\alpha^\dagger_\mu\right]=\omega_\mu\alpha^\dagger_\mu$, and the bosons $\alpha^\dagger_\mu$ satisfy the commutation relation $\left[\alpha^{\phantom{\dagger}}_\mu,\alpha^{\dagger}_\nu\right]=\delta_{\mu\nu}$. The Hamiltonian $\mathcal{H}^{(2)}$ becomes diagonal:
\begin{eqnarray}
\mathcal{H}^{(2)}=\sum^{60}_{\mu=1}\omega_{\mu}\left(\alpha^\dagger_\mu\alpha^{\phantom{\dagger}}_\mu+\frac{1}{2}\right)~.
\end{eqnarray}

\end{document}